# Data-driven Identification of Occupant Thermostat-Behavior Dynamics

Michael Kane[a,1], Kunind Sharma[a]

[a] Department of Civil and Environmental Engineering, Northeastern University, Boston, 02151, MA, USA

## ABSTRACT

Building occupant behavior drives significant differences in building energy use, even in automated buildings. Users' distrust in the automation causes them to override settings. This results in responses that fail to satisfy both the occupants' and/or the building automation's objectives. The transition toward grid-interactive efficient buildings will make this evermore important as complex building control systems optimize not only for comfort, but also changing electricity costs. This paper presents a data-driven approach to study thermal comfort behavior dynamics which are not captured by standard steady-state comfort models such as predicted mean vote.

The proposed model captures the time it takes for a user to override a thermostat setpoint change as a function of the manual setpoint change magnitude. The model was trained with the ecobee Donate Your Data dataset of 5 min. resolution data from 27,764 smart thermostats and occupancy sensors. The resulting population-level model shows that, on average, a 2°F override will occur after ~30 mins. and an 8°F override will occur in only ~15 mins., indicating the magnitude of discomfort as a key driver to the swiftness of an override. Such models could improve demand response programs through personalized controls.

[1] Corresponding author. Northeastern University, Department of Civil and Environmental Engineering, 360 Huntington Ave., Boston, MA 20115. Tel: +1 (617) 373-7193. E-mail address: mi.kane@northeastern.edu.

# HIGHLIGHTS

- Analyzed occupant behavior in a dataset of 27k thermostats and occupancy sensors
- Investigated the relationship between override features, behavior factors, and energy
- Compared this analysis' behavior factors to those of prior small-scale studies
- Calculated statistics of manual override timing and magnitude relationship

# ABBREVIATIONS

| | |
|---|---|
| DR | Demand Response |
| DRP | Demand Response Program |
| DoD | Degree of Discomfort |
| DyD | Donate Your Data |
| GEB | Grid-interactive Efficient Building |
| MSC | Manual Setpoint Change |
| PIR | Passive Infrared |
| PSC | Programmed Setpoint Change |
| SC | Setpoint Change |
| SES | Smart Energy Solutions |
| TTD | Time to Discomfort |

# 1 INTRODUCTION

The residential sector consumed 32% of the 101 quadrillion BTU of energy consumed in the US in 2018 [1]. The primary objective of this energy use was occupant comfort. Grid-interactive efficient buildings (GEBs) promise to turn these loads into assets for the grid through reducing, shedding, shifting, and modulating flexible building loads to balance renewable energy and help satisfy grid constraints [2].

Currently, utilities use demand response programs (DRPs) to send messages, remotely switch loads, or change thermostat setpoints to shed peak demand. For example, National Grid's Smart Energy Solutions (SES) pilot study in Worcester, MA, reduced peak loads by 27%–31% [3]. If enough occupants opt-out or override these automated changes, then the DRP could face performance penalties. There were $3.7M of such penalties in 2017 [4]. Utilities model the effect of weather and other factors to improve the accuracy of their commitment to reduce load on the grid. However, the same controls are sent to all participants, regardless of their sensitivity to setpoint changes, reduced quality of service, or economics. Such strategies resulted in up to 30% of SES pilot participants overriding or opting-out of an 8-hr event, and 10% for a 3-hr event [3].

In residential DRPs that switch-off HVAC systems during peak events or setback the thermostat (i.e., decrease/increase the setpoint heating/cooling season, respectively) [5,6], occupants may try to game the system by increasing the thermostat setpoint (i.e., in heating season) in anticipation of a demand response (DR) event [7]. Or, occupants who want to conserve energy may already have their thermostat set at the edge of their comfort zone, leading to extreme discomfort during DR events [8]. These user behaviors and subsequent overrides ultimately reduce the effectiveness of DRPs.

Personalized and predictive controls have been proposed to improve occupant satisfaction with building controls. However, in reviews by Mirakhorli and Dong [9] and Kim et al. [10] only two *dynamic* thermal comfort models are presented. In [11] a state-space model of thermal comfort is developed, but only validated with lab data from 13 subjects. A transient thermal comfort model is developed in [12] from 109 occupants, but for automotive applications. Real-time data on occupant comfort could improve occupant satisfaction through personalization. For example, Comfy® replaces the predicted mean vote estimator with real votes to determine thermostat setpoints for shared offices. Queries for real-time feedback should be limited though, as occupants have limited cognitive interest in reducing energy costs [13].

Better predictive models of thermal comfort behavior might arise from an understanding of the internal processes that govern thermal comfort *behavior*. Cognitive Dissonance, Theory of Planned Behavior, and Social Cognitive Theory have been used to model energy use [14], but these don't explicitly account for both choosing among options to improve comfort and physiological aspects. Occupant decision making can be modeled as an optimization problem with limited information and understanding of the system being optimized. Occupants' understanding of thermostats can be classified by two mental models: the incorrect *valve theory mental model*, which says that the larger the change to the thermostat's setpoint the faster the room will become comfortable, and the correct *feedback theory mental model*, which says that the room will reach the desired temperature as quickly as possible [15]. The occupant-building response thus consists of a building–physiology–cognition–decision feedback loop, such a modeling framework is presented in [16].

This paper aims to model this input-output building–physiology–cognition processes, and is organized as follows: Section 2 provides an introduction to the data, describes data conditioning and feature extraction techniques used, and outlines the proposed modeling framework. Section 3 provides results and the insights extracted from the data; and Section 4 contains conclusions and future work.

## 2   MATERIALS AND METHODS

The goal of the methods below is to quantitatively understand the dynamics of occupant comfort relevant to thermostat overrides. Foundational to such an analysis is the ecobee Donate Your Data (DyD) [17]. The data is preprocessed to correct for noise and sensor bias. Feature vectors such as the magnitude and timing of overrides are extracted and aligned with behavior models. The results are compared to other studies of occupant thermal comfort behavior.

## 2.1 DATASET

ecobee's customers can choose to opt-in to the DyD data sharing program. Their data is anonymized and disseminated to "help scientists advance the way to a sustainable future" [18]. Similar large datasets are available from Pecan Street Inc. [19] and private utility datasets [3,20], but ecobee's dataset is unique in that it includes occupant sensing and accurate indoor temperature.

The entire dataset used for this paper contains records from ~27k homes in ~28 countries, primarily in the US. The dataset for each user includes all data from when the user joined the data sharing program through September 2018. Part of the data is user-reported metadata including number of occupants, area of the house, age of the house, etc., and part is collected by ecobee thermostats (sampled every 5 min) [18]:

- time stamp
- cooling/heating setpoint
- runtime of each heating/cooling stage in the last 5 min.
- setpoint temperature
- event driven setpoint changes due to
    - demand response events
    - manual setpoint changes, may last for 2hr, 4hrs, until next event, or indefinitely
    - Awake/Away/Home/Sleep scheduled events
    - Smart Away/Smart Home/Smart Recovery event based on motion sensors or geofence
- PIR motion sensor data
- indoor temperature
- outdoor temperature

- and other data not relevant to this study

To avoid the more complex task of separating behavior of multiple occupants in one household, only the single-occupant households, as reported by the user (1,410 homes), are primarily considered. Most of these households are located in US (1,264 homes). To understand behavioral dynamics related to thermal comfort, only homes where occupants manually changed the instantaneous setpoint at least 10 times are considered. MATLAB is used to analyze the 26 GB of data on a 2.3 GHz machine with 20 cores and 200 GB memory.

## 2.2 DATA CONDITIONING

The data was converted from CSV to MATLAB files and pre-processed to identify the temperature unit (Celsius or Fahrenheit) used, remove setpoint sampling errors, and de-noise the occupancy sensor data. Each time the setpoint changed (SC), features where computed: the setpoint at previous setpoint change, the occupied time since last setpoint change, and the time to next setpoint change.

### 2.2.1 Identify User's Temperature Units (Celsius or Fahrenheit)

To understand a user's thermal discomfort and to identify data irregularities, it is important to know the setpoint units—Celsius or Fahrenheit; Celsius and Fahrenheit thermostats can be changed in $0.5°C$ and $1.0°F$ increments respectively. First, all setpoint changes are converted to $°C$. If these $°C$ increments modulo $0.5°C$ are less than 0.001, the users is assumed to use $°C$; if the $°F$ increments modulo $1.0°F$ are less than 0.001 the user is assumed to use $°F$. About 2.5% of the users could not be identified as using $°F$ or °C and their data are discarded.

### 2.2.2 Remove Setpoint Sampling Errors

Thermostat data is sampled every 5 mins by ecobee, and the occupant can manually change the setpoint at any time in that interval. Equation (1) shows how the sampled setpoint in the dataset $\tilde{x}_1$ is the time-average of the setpoint at the beginning $x_0$ and end $x_1$ of the 5 minute interval, where the switch occurred at $\Delta t$ minutes.

$$\tilde{x}_1 = \frac{\Delta t}{5} x_0 + \frac{5 - \Delta t}{5} x_1, \qquad (1)$$

For example, Fig. 1 shows sampled setpoint increments with non-standard values (i.e., not multiples of $0.5°C$) indicating a setpoint change occurred between 0-5 min. and 10-15 min. respectively. Without knowing the precise time that the setpoint change occurred, it is impossible to estimate the actual setpoint at 5.0 minutes[1]. Instead, it is assumed that $\Delta t = 2.5$ min, and (1) is solved for $x_1$ with observations of $x_0$ and $\tilde{x}_1$ and rounded to the nearest $1.0F$ or $0.5°C$ depending on the user, yielding the estimates shown.

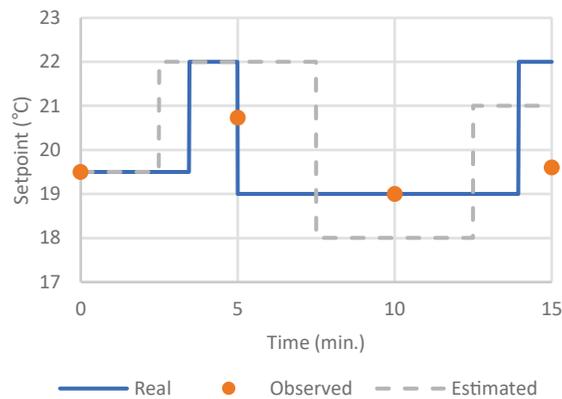

**Fig. 1:** ecobee's method of averaging temperatures for each data sample and the proposed estimation strategy

---

[1] If $\Delta t$ in (1) is uniformly distributed and $x_0$ and $\tilde{x}_1$ are observed, the expected value of $E[x_1] = \int_0^5 \frac{(5\tilde{x}_1 - \Delta t \cdot x_0)}{5 - \Delta t} (5) \, d\Delta t$ does not converge.

### 2.2.3 De-noise Occupancy Sensor Data

Focusing on the relationship between occupant comfort and thermostat setpoints, the data should be filtered to include only times when occupants where in the home as detected by any of the passive infrared (PIR) sensors on the ecobee thermostats and remote sensors in the home. The accuracy of PIR sensors are significantly biased, with false positive detection rates of only 3% but true positive detection rates of 33% [21]. To reduce the impact of false negatives, the data is filtered to 'fill in' short periods where the occupant may have been temporarily occluded from the sensor. The results in §3.1 show the effect of the length of this filtering window, from 0 to 120 minutes, on the average length of detected occupancy. This algorithm assumes the user did *not* leave and return to the house within this period, but rather was home the entire time. To reduce the impact of the accuracy bias, the analysis below relies only on the high accuracy occupied data; although, the length of occupied segments will increase with filter width as the short gap between segments are filled in, joining the segments.

## 2.3 EXTRACTING SETPOINT CHANGE (SC) FEATURES

Each time the setpoint changed (SC), the datapoint is tagged as a programmed setpoint change (PSC) (e.g., due to a schedule or based on occupancy) or a manual setpoint change (MSC) (e.g., the occupant changed the setpoint to make themselves more comfortable or save energy). As mentioned in §2.2.3, only the MSC points where occupancy is continuously detected (for a period of less than 2 hours) in the filtered data following the previous SC were considered for analysis. Experiments by Kolarik et al. have shown that after two hours, occupants are fully adapted to the environment, thereby operating in steady steady-state with negligible dynamics [22].

Initial analysis of the data (§3.5) showed that ~60% of MSCs resulted in more energy consumption (e.g., increased/decreased the setpoint during heating/cooling, respectively), justifying the simplifying assumption of this study that during occupied periods, MSCs are caused by occupants trying to improve their comfort. This framing yields two important features from each MSC: the degree of discomfort (DoD), i.e., the magnitude of the MSC; and the time to discomfort (TTD), i.e., the time elapsed following the previous SC.

## 2.4 BEHAVIOR MODELS

Each MSC is a consequence of a number of interdependencies between the house, the HVAC system, and the user [16,23] that form the feedback loop shown in Fig. 2. **Building Physics**: HVAC equipment size, ventilation/airspeed, infiltration, solar radiation, etc. drive indoor temperature dynamics [24]. **User physiology**: skin temperature and internal temperature change as a function of time, temperature, clothing, metabolism, etc. [25]. **Cognition**: internal cues driven by body temperature are combined with perceived barriers, observed behavior, and other inputs into cognitive states (e.g., level of self-efficacy, cue to action, and behavior) that integrate and trigger actions [16]. **Action**: to achieve comfort, users may change clothing, open windows, or modify the thermostat. If the later, what they change it to will depend on their understanding of how thermostats work, i.e. their *mental model*. A *valve theory* mental model incorrectly assumes higher MSC magnitudes will results in faster changes in room temperature, a vestige of radiators that worked this way when the steam valve was opened. Previous research has shown that ~$\frac{1}{3}^{rd}$ of the population incorrectly operates thermostats in this manner, while the remaining ~$\frac{2}{3}^{rds}$ utilize the correct *feedback theory* mental model of thermostats [15]. **Thermostat**: Most residential HVAC thermostats maintain the temperature within the setpoint deadband by switching equipment on/off.

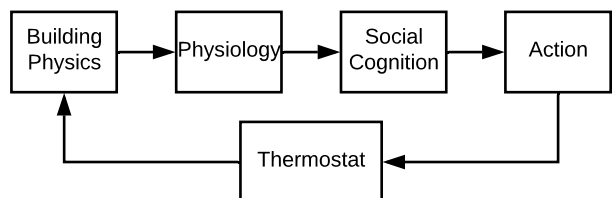

**Fig. 2:** Human-in-the-loop model of a building

Without additional sensors or user surveys, this paper studies only the input-output behavior of the building-physiology-cognition-action process. However, mental models could be inferred from the data if users "set it and forget it", indicating *feedback theory* or "fiddle" with the setpoint resulting in overshoot behavior indicative of *valve theory* mental models or a significant misunderstanding of what temperature they'd feel comfortable.

## 2.5 ANALYSIS APPROACH

The methods are applied to the data in the order presented above. First, each users' temperature unit is identified and the setpoint data de-averaged. The relationship between occupancy filter length (no filter, 15 min., 20 min., 30 min., and 120 min.) on the length of occupied periods is then studied to identify a filter length that balances filling in false-negatives in the raw data with the potential for false-positives in the filtered data (§3.1). Although, without ground-truth occupancy data, this can only be done heuristically. Even after choosing a filter length, the impact of the decision on each later analysis should be revisited.

After identifying each SC and then MSC, the time of day and duration until the next SC for each MSC will yield insights into schedule-based causes and energy impacts of user overrides (§3.2). Studying the statistics of the TTDs for all the MSCs will show how dynamic user behavior is, and potentially question the efficacy of static ASHRAE comfort models for GEBs (§3.3).

The events (e.g., away, awake, home, etc.) that lead to the SC preceding each MSC will yield insights into the relationship between the user and the thermostat, and potentially user mental models (§3.4). The DoD for each MSC will yield further insights into the energy impacts of users' decisions based on the relative number of MSCs where setpoints are increased or decreased during heating or cooling.

The core analysis of this paper is the relationship between TTD and DoD and statistical quantification of the "time to discomfort" given the "degree of discomfort" (§3.5).

# 3 RESULTS AND DISCUSSION

## 3.1 PRE-PROCESSING

Fig. 3 shows the effect of the filter length on the detected periods of occupancy. The large number of short periods in the raw data indicates many false negatives that split up extended periods when the occupants are home (e.g., a few hours at home after work and before bed). As the filter length increases to 10 min., 20 min., and then 30 min., more and more of these broken apart long periods are stitched together, increasing the number of longer occupied periods suitable for comfort studies by a half an order of magnitude.

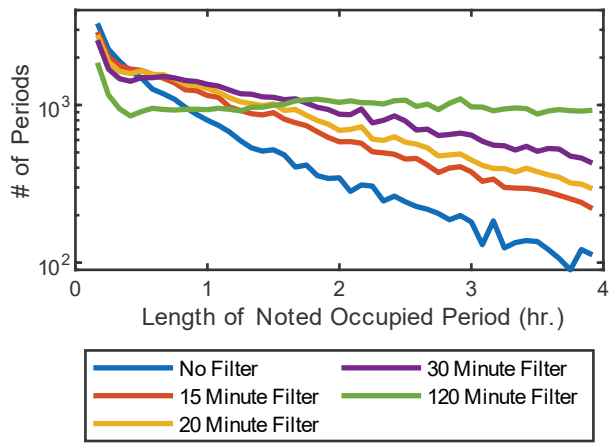

**Fig. 3:** Comparison of different filters on the occupancy data

The 120 min. filter decreases the number of noted occupied periods overall but yields more data points with occupied periods greater than 3.5 hrs., which may not be of use as most of the MSCs dynamics are assumed to take place within 2 hrs. of a triggering event. Moreover, while the 120 min. filter does convert false negatives to true positives, a lot of true negatives may also be converted to false positives. In other words, even when the person was not home for most of a 2 hr. period, the entire 2 hrs. were assumed to be occupied, potentially corrupting the data of thermal comfort behavior governed by outdoor temperatures.

In lieu of ground truth or alternate occupancy data, the 30 min. filter seems to serve as the middle ground for false positives and false negatives and has also been used in prior studies [21].

## 3.2 TIME OF DAY AND DURATION OF OVERRIDES

Fig. 4 depicts the MSCs throughout the day for single-occupant households and shows distinct periods of frequent overrides. It is helpful to frame this figure with approximations of typical setback schedules programmed into the thermostat: e.g., *sleep* from 10 PM–6 AM, *wake-up* from 6 AM–9 AM, *away* from 9 AM–6 PM, and *evening home* from 6 PM–10 PM during weekdays, and the away period removed during the weekends. The weekday spike in overrides mid-morning could then be early birds setting back their thermostat before leaving early for work or occupants spending the day at home and overriding the away period. The other large spike in overrides occurs in the evening, and may be switching to sleep mode early, or night owls extending their evening comfort zone into midnight. The average occupant manually changes the thermostat 0.9 times per day.

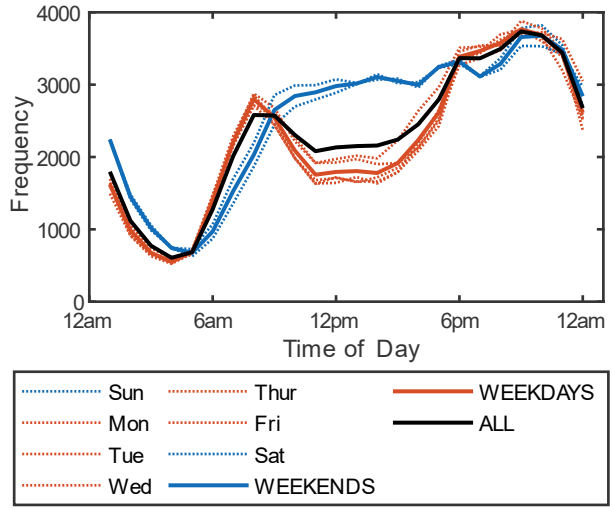

Fig. 4: Trends in manual setpoint changes during the day
(single-occupant households; n=1,401)

The number and duration of MSCs with respect to the time of day are plotted in a 2-D histogram in Fig. 5, and may serve as a proxy for the energy impact of overrides. The MSCs with the longest period of impact were the night owls, who's setbacks lasted 6-12 hrs. overnight. The occupants staying home in the morning also had extensive impacts, lasting 6-12 hrs. throughout the day. The large number of MSCs in the early evening lasted only 1.5-3 hrs., indicative of a dissatisfaction in the automation's ability to set comfortable temperatures during this post-work, pre-sleep period.

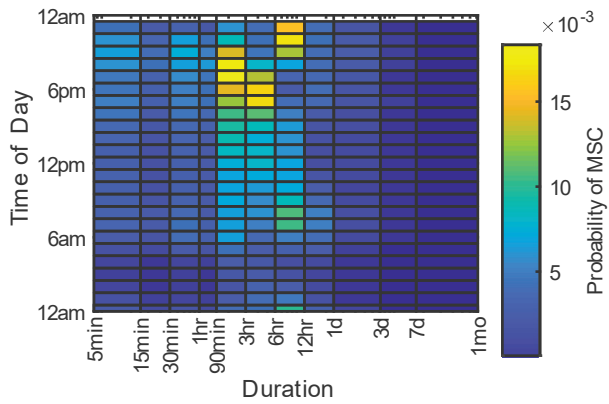

Fig. 5: Number and duration of manual setpoint changes with
respect to the time of day (single-occupant households)

## 3.3 TIME TO DISCOMFORT

To understand occupant behavior patterns that *lead to* MSCs, occupants' time to discomfort (TTD) is calculated. I.e., Fig. 6 statistically shows how long it takes for occupants, already in the space prior to the SC, to feel uncomfortable and make an MSC.

The majority of MSCs occur within the first 10 mins of an SC, where the occupant immediately notices or predicts the setpoint change. The more interesting cases are the increasing TTDs peaking around 15-20 mins and exponentially trailing off. The plot is limited to 4 hrs. because the data after that is minimal, having a probability density of less than 0.01.

On average, occupants are in the space for 55 mins. prior to overriding the system; however, if the first peak is removed, the average shifts to 72 mins. The median time to discomfort of 15–20 min could be influenced by any of the factors of the building-physiology-cognition-action process and likely varies from person to person and building to building.

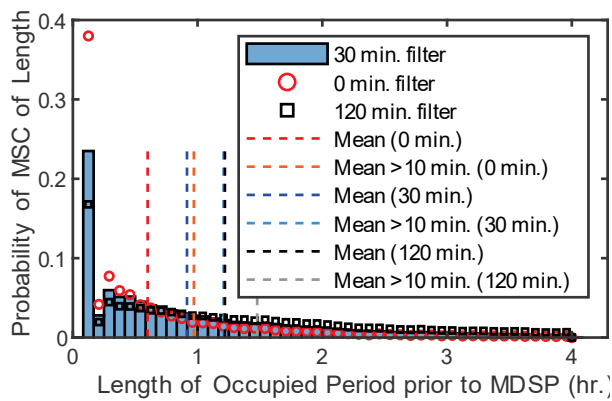

**Fig. 6:** Time from setpoint change to manual setpoint change (occupied; 30, 0, 120 min.)

The distributions of TTD for various occupancy filter lengths are also indicated in the figure. As expected, the unfiltered data shows more TTDs around the 15 min. median, while the 120 min. filter decreases the prevalence of these short TTD and increases the prevalence of longer (>1 hr.) TTDs.

Fig. 7 shows the distribution of events causing the SC prior to each MSC. Event types are defined in §2.1. Over half of all MSC were overriding a previous setpoint by that user, not overriding the automation. It is possible that many of these overrides, some corresponding to the median in Fig. 6, could be due to occupant's incorrect *valve theory* mental model. From an automation standpoint, the users are not satisfied with the scheduled-based *Home* and *Sleep* PSCs, preceding ~30% of all MSCs. Users seem to place more trust in the *Smart* features, rarely overriding them; although, this could be due to the low prevalence of these events in the dataset. The dataset has too few *Demand Response* events to draw meaningful insights currently.

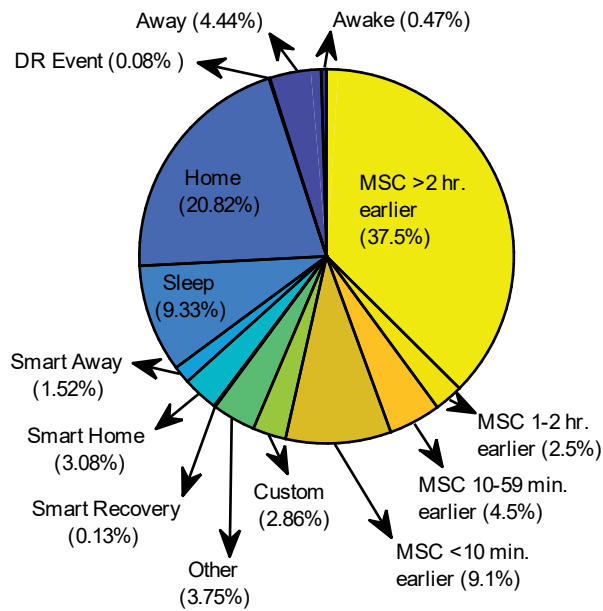

**Fig. 7:** Event immediately prior to a manual setpoint change (single-occupant households)

## 3.4 MENTAL MODELS

Approximately 20% of users override *more than once in a row* after an initial MSC, exhibiting traits related to the valve theory mental model. The 52% of users who made an override *only once* after an initial MSC seem to be trying to follow the feedback theory mental model but behave non-optimally due to an apparent inability to estimate their own comfort zone. The remaining ~28% of users *make no additional MSCs* within 1 hr. after an MSC and likely utilize a valid feedback theory mental model. This aligns well with prior research showing ~30% valve theory and ~70% feedback theory [15].

## 3.5 IMPACT OF MANUAL SETPOINT CHANGES

Fig. 8 shows the distribution of MSCs' *degree of discomfort (DoD)* (assuming the occupant is comfortable after the override) made in heating and cooling seasons with respect to the indoor temperature. The median setpoint change is ~$2°F$ during the cooling season with an indoor temperature ~74°F, and a $2°F$ increase during the heating season with an indoor temperature ~68°F. In the cooling season, 58% of these MSC increase energy use, while 57% increase the energy use in the heating season. See [17] for a thorough coverage of indoor temperatures and energy impacts in the DyD dataset.

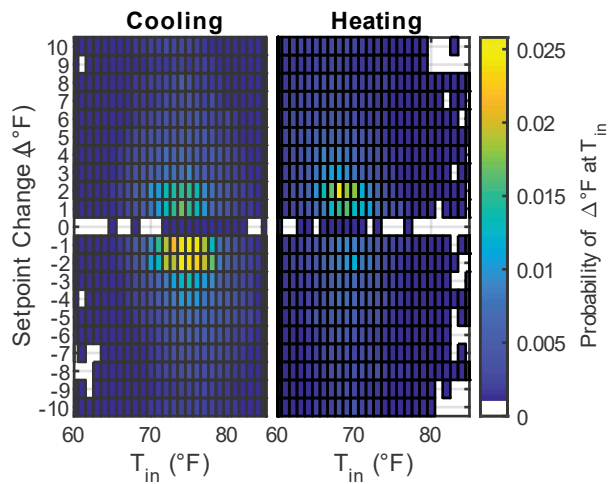

**Fig. 8:** Manual setpoint changes by season (single-occupant households)

## 3.6 BEHAVIOR DYNAMICS

Variables that influence thermal comfort include TTD, how long an occupant (who is at home) takes to feel uncomfortable after any kind of a setpoint change (a PSC or an MSC) and makes an MSC; and DoD, the change in temperature of the MSC. The relationships between these variables could guide DRPs that setback thermostats to temperatures that would minimize overrides for the desired period, thus increasing DRP reliability.

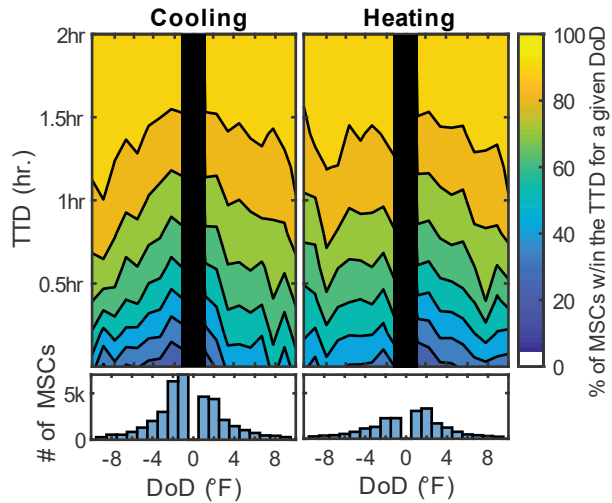

**Fig. 9:** Degree of discomfort vs Time to discomfort; during occupied periods (Single-occupant households; 30 min. filter)

Fig. 9 shows the decreasing relationship between TTD and increasing DoD. In the cooling season, ~28% of MSCs of -2°F took place within 10 mins., 50% took place within 30 mins., and 60% took place within 40 mins. Equation (2) and (3) are inverse exponential fits to these plots for 50% of the population considering negative DoDs during cooling season, and positive during heating season, respectively (i.e., MSCs that would result in an increase in energy use).

$$TTD = 0.5368 \cdot e^{-0.083 \cdot DOD} \qquad (2)$$
$$TTD = 0.5804 \cdot e^{-0.074 \cdot DOD} \qquad (3)$$

The close similarity between energy intensive behavior dynamics in heating and cooling seasons contrasts with the difference in energy intensive behavior versus energy saving behavior. Energy saving MSCs (i.e., those triggered by positive DoDs during cooling season, and negative DoDs during heating season) have significantly faster TTDs compared to energy intensive MSCs. These results reinforce observations from [22] that increasing temperatures are sensed at different rates than decreasing temperatures, yet the results presented here suggest that difference could be seasonally affected.

It should be noted that the pre-processing done for this analysis, using a 30 min. filter, plays an important role in determining the TTD. Fig. 10 shows the relationship between DoD and TTD for different occupancy filters. As expected, the TTD increases as the occupancy filter length increases, due to more data of longer occupied periods entering the MSC dataset used for analysis. Moreover, the number of MSCs continuously occupied following the previous SC (within 2 hrs.) increases with the length of the filter. E.g., the number of 1–2°F MSCs for no filter, 30 min. filter, and 120 min. filter are ~4k, 7k, and 10k respectively.

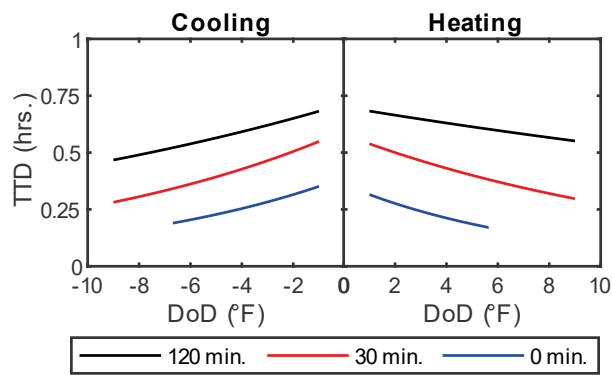

**Fig. 10:** Effect occupancy filter on the relationship between degree of discomfort and time to discomfort

## 3.7 ENTIRE DATASET

The results above are computed using the subset of the DyD dataset self-reported as having only one occupant. This provided insights into the behavior of individuals, removing conflating factors due to social barriers to comfort. Multiple-occupant homes will lead to long occupancy periods because different people move in and out of the home at different times. The dynamics of thermal comfort will then be based on data for different occupants with entirely different metabolisms, clothing, behavior patterns, etc. To understand the impact of these social factors, the methods above are applied to the entire DyD dataset, which consists of ~27k homes (including homes that users report as "single occupant" and "multiple occupant," as well as no reported category).

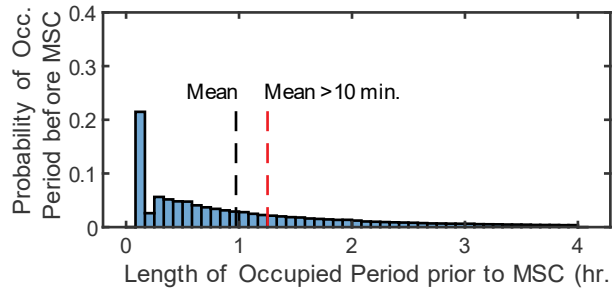

**Fig. 11:** Entire dataset: Time from setpoint change to manual setpoint change (occupied; 30 min. filter)

Comparing Fig. 11, the distribution of TTD for the entire dataset, with Fig. 6, the distribution for single-occupant households only, shows similar thermostat behavior regardless of social factors. The largest number of MSCs occur within 10 min. of the previous SC (24% and 21% for single occupancy and all data, respectively), and the next median TTD occurs within 15-20 min. The mean of the data after 10 min. is essentially unchanged. Likewise, comparing the event leading to the MSCs for single-occupant households (i.e., Fig. 7) and all data would show similar miniscule changes.

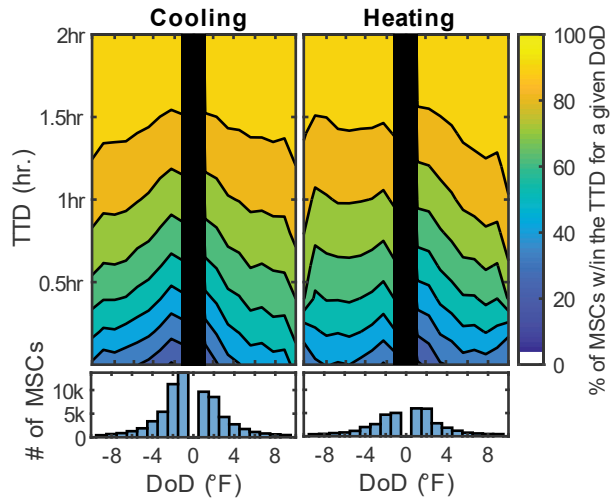

**Fig. 12:** Entire dataset: Degree of discomfort vs Time to discomfort

Similar to Fig. 9, the contour lines in Fig. 12 shows the relationship between TTD and DoD. The contours are smoother for Fig. 12 using the entire dataset, due to the availability of more MSC data points. Table 1 compares key values between Fig. 9 and Fig. 12, namely the period after which 50% of the population will have become uncomfortable and trigger an MSC (i.e., TTD-50%) for low and high magnitudes of discomfort (i.e., DoD). The similarity in these numbers seems to indicate that social factors have little appreciable impact on thermal comfort behavior dynamics as observed through thermostat changes.

**Table 1**

*COMPARISON OF DOD VS. TTD-50% FOR SINGLE VS ALL OCCUPANCY DATA*

| Heating | $2°F$ DoD | $8°F$ DoD |
|---|---|---|
| Single occ. | 28 min. | 15 min |
| All data. | 30 min. | 17 min. |
| Cooling | $-2°F$ DoD | $-8°F$ DoD |
| Single occ. | 28 min. | 15 min. |
| All data. | 29 min. | 18 min |

# 4 CONCLUSIONS

This paper analyzed ecobee's "Donate Your Data" dataset of ~27k smart thermostats to illustrate and understand the dynamic nature of occupant thermal comfort behavior, especially relating to thermostat use. The analysis focused on the subset of data self-declared as single occupant homes to best understand individual behavior; however, extending the methods to the entire dataset did not substantially change the results. Each manual setpoint change (MSC) in the data was identified and those which were occupied continuously following the previous setpoint change (SC) were considered for analysis. The analysis then assumed that the previous SC led to the occupants' discomfort and the MSC set the thermostat to a comfortable temperature.

Programmed setpoint changes (PSCs) accounted for less than half the events preceding the MSCs, showing that many occupants have not programmed their thermostat or are overriding decisions they previously made while continuously occupying the space. This later behavior, an apparent iterative process of finding comfortable setpoints, supports the finding in other literature that ~30% of occupants have an incorrect *valve theory* mental model of how a thermostat works [15]. About 60% of the MSCs in both heating and cooling modes resulted in increased energy consumption.

The principal result of this analysis captures the input-output relationship of the building-physiology-cognition-action process that leads to an MSC correcting discomfort caused by a previous SC. An understanding of this process will be important for designing adaptive and personalized grid interactive efficient buildings (GEBs). The time it takes for an occupant to notice and correct a setpoint is often nearly instantaneous (if they predict the thermostat schedule or the hear/see the thermostat change), but the majority of MSCs occur some occupied period after the SC. This time to discomfort (TTD) is inversely exponentially related to the degree of discomfort (DoD): in both heating and cooling mode, 50% of occupants will take less than ~30 min. to correct an SC that makes them $2°F$ too cool/warm, but less than ~15 min. for being $8°F$ too cool/warm respectively.

These results could be immediately useful to demand response programs (DRPs) to choose the duration and magnitude of connected thermostats setbacks that will result in statistically acceptable load curtailment accuracy, by estimating not only override rates, but timing as well. Further accuracy improvements could be made by creating personalized models of thermal comfort behavior dynamics for deploying personalized DRPs that maximize curtailment and reliability. Further research is required to truly understand these phenomena however, as this behavior is an outcome of a complex multi-disciplinary feedback loop encompassing building physics, human physiology, cognition, and decision making.

# ACKNOWLEDGEMENTS

The authors would like to acknowledge their gratitude to ecobee's Social Impact Department for curating and sharing the Donate Your Data dataset that is foundational to this research. Other researchers may request no-cost access to this data by visiting https://www.ecobee.com/donateyourdata/signup/

The authors would also like to acknowledge their appreciation for the editing services provided by Susan Matheson for the preparation of this manuscript.

This research did not receive any specific grant from funding agencies in the public, commercial, or not-for-profit sectors.